\documentclass{jfm}

\usepackage[usenames,dvipsnames]{xcolor}
\usepackage{graphicx}
\usepackage{amsmath}
\usepackage{amssymb}
\usepackage{amsfonts}
\usepackage{natbib}
\usepackage{psfrag}
\usepackage{wasysym}
\usepackage{float}
\usepackage{epstopdf}
\usepackage{color}
\usepackage[normalem]{ulem}
\usepackage{textcomp}

\renewcommand{\d}[2]{\frac{\id #1}{\id #2}} %< for derivatives
\newcommand{\overbar}[1]{\mkern 1.5mu\overline{\mkern-1.5mu#1\mkern-1.5mu}\mkern 1.5mu}
\newcommand{\id}{\mathrm{d}} %for integral d...
\newcommand{\ii}{\mathrm{i}}
\newcommand{\e}{\mathrm{e}}
\newcommand{\T}{\mathcal{T}}

\shorttitle{Viscoelastic lubrication}
\shortauthor{A. Pandey and others}

\title{Lubrication of soft viscoelastic solids}

\author{A. Pandey\aff{1}, S. Karpitschka\aff{1}, C. H. Venner\aff{2} \and J. H. Snoeijer\aff{1,3}}

\affiliation{\aff{1}Physics of Fluids Group, Faculty of Science and Technology, University of Twente, P.O. Box 217, 7500 AE Enschede,The Netherlands
\aff{2}Faculty of Engineering Technology, Engineering Fluid Dynamics, University of Twente, P.O. Box 217, 7500 AE Enschede, The Netherlands
\aff{3}Department of Applied Physics, Eindhoven University of Technology,
P.O. Box 513, 5600MB Eindhoven, The Netherlands }

\begin{document}
\maketitle
\begin{abstract}
Lubrication flows appear in many applications in engineering, biophysics, and in nature. Separation of surfaces and minimisation of friction and wear is achieved when the lubrication fluid builds up a lift force. In this paper we analyse soft lubricated contacts by treating the solid walls as viscoelastic: soft materials are typically not purely elastic, but dissipate energy under dynamical loading conditions. We present a method for viscoelastic lubrication and focus on three canonical examples, namely Kelvin-Voigt-, Standard Linear-, and Power Law-rheology. It is shown how the solid viscoelasticity affects the lubrication process when the timescale of loading becomes comparable to the rheological timescale. We derive asymptotic relations between lift force and sliding velocity, which give scaling laws that inherit a signature of the rheology. In all cases the lift is found to decrease with respect to purely elastic systems.
\end{abstract}

%%%%%%%%%%%%%%%%%%%%%%%%%%%%%%%%%%%%%%%%%%%%%%%%%%%%%%%%%%%%%%%%%%%%%%%%%%%%%%%
\section{Introduction}

The `art' of lubrication by thin liquid layers is known since ancient times~\citep[]{dowson}, permitting motion between adjacent solid surfaces at low friction and wear. Lubrication is of paramount importance to the safe, reliable, and controlled operation of many key elements in engineering applications ranging from very large scale (planes, windturbines) to microfluidic devices. Synovial joints in mammals are the archetype of this mechanism in nature. From a theoretical point of view, the flow of a liquid within a narrow gap can be described by the lubrication approximation of Stokes equations, first developed by Reynolds more than a century ago \citep[]{Reynolds1886}. Since then, lubrication theory has been used to understand a wide range of phenomena like moving bubbles in a tube \citep[]{Bretherton1961}, motion of red blood cells in capillaries \citep[]{Fitz1969, Secomb1986, Feng2000}, bio-mechanics of articular cartilage \citep[]{Hou1992, Mow1993} or the physics of `Kugel fountain'~\citep[]{SnoeijerAJP14}, to name a few examples. 

Due to the reversibility of Stokes flow in a lubricating layer between rigid solids, the lift force 
\begin{equation}
L=\int p\, \id A,
\label{L}
\end{equation}
between the two sliding or rotating bodies vanishes: in a non-cavitating liquid a lubricated contact could not support any load if the bodies were entirely rigid. However, usually the counter-moving bodies are deformable. Importantly, the deformation was found to break the reversibility of Stokes flow in the lubricating layer, which generates a lift force $L>0$ between the bodies \citep[]{Hooke1972,Bissett1989,Sekimoto1993,Snoeijer2013}. Not the least motivated by novel biological or bio-inspired engineering applications, this problem has been addressed on many occasions in the last decade. Subjects range from `soft lubrication' \citep[]{Martin2002, Skotheim2004, Skotheim2005a}, motion of lubricated eyelid wiper \citep[]{Jones2008}, or sticking of particles on lubricated compliant substrates \citep[]{Mani2012}, to translating, spinning particles near soft boundaries \citep[]{Urzay2007, Salez2015}. 

In most studies, the deformation of elastic materials was assumed to adapt instantaneously to the stresses at their boundaries. In practice, however, most soft materials like gels, elastomers or cartilage behave viscoelastically: due to dissipation, their relaxation behavior is time dependent, and deformation requires finite time to adapt to changes in the loading. For example, recent experiments show the viscoelastic nature of articular cartilage during osteoarthritis \citep[]{Trickery2000, Desrochers2012}. Hence the coupling between lubrication pressure and viscoelastic deformation is crucial in understanding these systems \citep[]{1997,Scaraggi2014}.

\begin{figure}
	\begin{center}%
		\includegraphics[width=135mm]{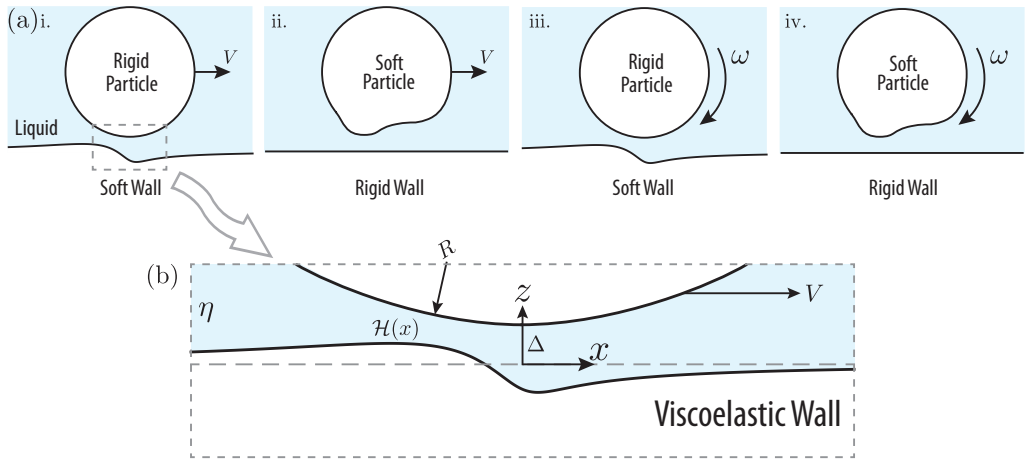}%
	\end{center}%
	\vspace{5mm}
\caption{(a) Four typical geometries of two-dimensional soft lubrication. The arrows show the direction of steady motion of the cylinder. If the deformed material is purely elastic in nature all four has same steady state response when studied in a comoving frame attached with the cylinder. Viscoelasticity breaks the symmetry of the problem: the material coordinates in cases (i., iv.) are exposed to a dynamical loading, and probe the time-dependent rheology of the solid. (b) Schematic of the problem studied here: a steadily moving rigid cylinder close to a viscoelastic wall, corresponding to case (i.). Note that case (iv.) can be recovered by setting $V\rightarrow \omega R$.}
	\label{fig:1}
\end{figure}

In this paper we investigate how the lubrication is affected by viscoelasticity of the deformable boundaries. Figure~\ref{fig:1}(a) shows four principle configurations in steady soft lubrication in a two-dimensional configuration: translation with constant velocity $V$ (cases (i., ii.)) or rotation with constant frequency $\omega$ (cases (iii., iv.)) of cylinders near a wall. Either the wall (i., iii.) or the cylinder (ii., iv.) is assumed to be soft. When the soft material is assumed to be perfectly elastic, adapting instantaneously to changes in loading, all four cases are equivalent. At constant separation distance, the lift force and the velocity are then related as $L \sim V^2$ when deformation is small \citep[]{Skotheim2004}, and as $L \sim V^3$ when deformations are large \citep[]{Bissett1989,Snoeijer2013}.

Importantly, viscoelasticity breaks the symmetry between the various cases in figure~\ref{fig:1}(a). In cases (ii., iii.), the deformations are stationary relative to the material coordinates of the solid body. In stationary conditions, one only probes the long-time relaxation and the response is purely elastic. Contrarily to this, in cases (i., iv.), the deformation continuously propagates relative to the cylinder: material coordinates are exposed to a dynamical loading. This continuous reconfiguration of the viscoelastic material dissipates energy and probes the time-dependent rheology of the solid. The key question addressed here is how the viscoelastic rheology affects the generated lift force.

The paper is organised as follows. In section~\ref{sec:formulation} we formulate the problem based on linear viscoelastic response and discuss the solution strategy. Section~\ref{sec:results} presents the deformation profiles for three canonical examples, namely Kelvin-Voigt-, Standard Linear-, and Power Law-rheology. We show viscoelasticity affects the lift force in these cases, by a combination of numerical solution and asymptotic analysis. The paper concludes with a discussion in section~\ref{sec:discussion}.

%%%%%%%%%%%%%%%%%%%%%%%%%%%%%%%%%%%%%%%%%%%%%%%%%%%%%%%%%%%%%%%%%%%%%%%%%%%%%%%
\section{Formulation}\label{sec:formulation}

We focus on the geometry sketched in figure~\ref{fig:1}(b), where the cylinder is treated rigid and the lower boundary as viscoelastic. We restrict the analysis to small deformations, in which case the analysis equally applies to case (iv.), using the connection $V\rightarrow \omega R$. Below we  first formulate the problem, introduce dimensionless variables and subsequently explain the solution strategy to solve the viscoelastic lubrication problem. 

\subsection{Lubrication equation and Viscoelastic deformation}

A rigid cylinder of radius $R$ moves with a constant velocity $V$ within a fluid of dynamic viscosity $\eta$. The minimum distance between the cylinder and the undeformed wall is $\Delta$ (see figure~\ref{fig:1}(b)). In the limit $\Delta\ll R$, the cylinder is assumed to be parabolic and the gap profile is given by $h_0(x)=\Delta+x^2/2R= \Delta (1 + x^2/2R\Delta)$. Hence, the characteristic length of the contact zone becomes

\begin{equation}
\ell=\sqrt{2\Delta R}.
\end{equation}
The motion of the cylinder creates a lubrication pressure in the gap and this in turn deforms the wall. This deformation is characterized by $\mathcal{H}(x)$. The deformed gap profile is $h(x)=h_0(x)-\mathcal{H}(x)$. The profile of the thin gap $h(x)$ and the fluid pressure $p(x)$ in the gap are the unknowns of this coupled problem. They are related by two equations: the steady state hydrodynamic lubrication equation, and the relation between load and deformation of an viscoelastic half space. 

We first compute the deformation by treating the lubrication pressure as a traction acting on a semi-infinite viscoelastic solid layer. The stress in an incompressible (Poisson ratio $\nu=1/2$), linear, viscoelastic material under dynamic strain is 
\begin{equation}
\sigma_{ij}(x,z,t)=\int_{-\infty}^t\Psi(t-t')\dot{\epsilon}_{ij}(x,z,t')\id t'-\Pi(x,z,t)\delta_{ij},
\label{sigt1}
\end{equation}
where $\sigma_{ij}$ is the stress tensor, $\epsilon_{ij}$ is the strain tensor, $\Psi(t)$ is the shear relaxation modulus and $\Pi$ is the isotropic part of the stress tensor. The overhead dot represents a time derivative. Throughout we will assume plane strain conditions. For the case of inertia-free dynamics, mechanical equilibrium is defined by $\sigma_{ij,j}=0$. We apply a Fourier transform in time (defined as $\widehat{f}(\omega)=\int_{-\infty}^{\infty}f(t)\e^{-\ii\omega t}\id t$) to get
\begin{subequations}
\label{eqf}
\begin{align}
&\widehat{\sigma}_{ij}(x,z,\omega)=\mu(\omega)\widehat{\epsilon}_{ij}(x,z,\omega),\\
&\widehat{\sigma}_{ij,j}(x,z,\omega)=0
\end{align}
\end{subequations}
for the stress-strain relation and the equilibrium condition, respectively. $\mu(\omega)$ is the complex shear modulus of the material and is given by
\begin{equation}
\mu(\omega)=\ii\omega\int_0^{\infty}\Psi(t)\e^{-\ii\omega t}\id t=G'(\omega)+\ii G''(\omega),
 \label{SnLm}
 \end{equation}
where $G'(\omega)$ and $G''(\omega)$ are the storage and loss moduli. \eqref{eqf} can be solved for the surface profile $\mathcal{H}$ by a Green's function approach. Recently, a similar formulation has been used to study dynamic deformation of viscoelastic substrate under a moving contact line \citep[]{Karpitschka15}. 
 For an arbitrary, dynamic traction $\widehat{p}(x,\omega)$, applied at the top surface of the wall ($z=0$), the deformation is given by 
 \begin{equation}
 \widehat{\mathcal{H}}(x,\omega)=\int_{-\infty}^{\infty}\widehat{p}(x',\omega) \frac{\mathcal{K}(x-x')}{\mu(\omega)}\id x'
 \label{hfw}
 \end{equation} where $\mathcal{K}(x)$ is the elastic Green's function. Taking a Fourier transform of~\eqref{hfw} in space (defined as $\widetilde{f}(q)=\int_{-\infty}^{\infty}f(x)\e^{-\ii q x}\id x$), we get 
 \begin{equation}
 \widehat{\widetilde{\mathcal{H}}}(q,\omega)=\widehat{\widetilde{p}}(q,\omega) \frac{\widetilde{\mathcal{K}}(q)}{\mu(\omega)}.
\label{hfqw}
\end{equation} The Green's function for an elastic half-space is $\mathcal{K}(x)=\frac{\log|x|}{2\upi}$, or, in Fourier space, $\widetilde{\mathcal{K}}(q)=-\frac{1}{2|q|}$ \citep[]{Johnson1987contact}.  

In the present problem the cylinder moves at a constant velocity, so that the dynamical loading has the form of a traveling pressure wave $p(x-Vt)$. This simple form of the temporal loading enables detailed analysis taking into account the full history-dependent response. Taking Fourier transforms of $p(x-Vt)$ with respect to space and time, we reach $\widehat{\widetilde{p}}(q,\omega)=2\upi\widetilde{p}(q)\udelta(\omega+Vq)$, where $\udelta(\omega)$ is the Delta function. Then~\eqref{hfqw} simplifies to 
\begin{equation}
 \widehat{\widetilde{\mathcal{H}}}(q,\omega)=-\frac{\upi\widetilde{p}(q)}{|q|}\frac{\udelta(\omega+Vq)}{\mu(\omega)}.
 \label{hfqw1}
 \end{equation}
Now we take a backward transform from $\omega$ to $t$ (defined as $f(t)=\int_{-\infty}^{\infty}\widehat{f}(\omega)\e^{\ii\omega t}\frac{\id \omega}{2\pi}$),
\begin{equation}
\widetilde{\mathcal{H}}(q,t) = -\frac{\widetilde{p}(q)}{2|q|}\frac{\mathrm{e}^{-\ii V q t}}{\mu(-Vq)}.
\end{equation}
The only remaining time dependence is the phase factor $\e^{-\ii V q t}$, which describes the translational motion of the cylinder. Hence, in the \emph{comoving} frame that travels with the cylinder, the profile becomes 
\begin{equation}
\mathcal{H}(x)=\int_{-\infty}^{\infty}-\frac{\widetilde{p}(q)}{2|q|\mu(-Vq)}\e^{\ii qx}{\id q\over2\upi}.
\label{hreal}
\end{equation}
This is the first key equation of the problem, relating the lubrication pressure in the narrow gap to the deformation of the viscoelastic wall. 

The second equation is obtained by the steady-state lubrication equation, describing the Stokes flow in the narrow gap. In the frame comoving with the cylinder this reads 

\begin{equation}
\d{}{x}\left[\frac{1}{6\eta}h^3\d{p}{x}+ V h\right]=0,
\label{dlub}
\end{equation}
where remind that $h(x)=h_0(x) - \mathcal{H}(x)$ is the thickness of the liquid layer.~\eqref{hreal} and~\eqref{dlub} form a coupled set of equations that constitute the viscoelastic lubrication problem. For $\mu$=constant, this is the same set of equations as for ``classical'' 2D elastohydrodynamics \citep[]{Bissett1989,venner2000multi,Snoeijer2013}.
%%%%%%%%%%%%%%%%%%%%%%%%%%%%%%%%%%%%%%%%%%%%%%%%%%%%%%%%%%%%%%%%%%%%%%%%%%%%%%
\subsection{Non-dimensionalisation}\label{ndvs}
We use the contact length $\ell$ as the horizontal length scale and the gap height $\Delta$ as the vertical length scale. The lubrication pressure then scales as
\begin{equation}
\mathrm{P}^*=\eta\ell V/\Delta^2.
\end{equation}
The scale of the deformation induced by this lubrication pressure is given by
\begin{equation}
\mathcal{H}^*=\mathrm{P}^*\ell/ 2G=\eta VR/G\Delta,
\end{equation}
where $G$ is the \emph{static} shear modulus of the viscoelastic material, defined as $G'(\omega=0)$. 
Hence, it is natural to introduce the first dimensionless parameter of the problem as
\begin{equation}
\beta \equiv \frac{\mathcal{H}^*}{\Delta}=\frac{\eta V R}{G\Delta^2},
\label{beta}
\end{equation} 
which is the ratio of elastic deformation and typical gap size. In this paper we solve the coupled equations in the limit where $\mathcal{H}^*$ is small compared to $\Delta$, i.e. $\beta \ll 1$.

In contrast to the purely elastic case, the viscoelastic wall exhibits a relaxation timescale~$\tau$. This timescale needs to be compared to that of the dynamical loading due to the lubrication pressure. This pressure evolves on a timescale $\tau_{\rm p}=\ell/V$, which can be seen as the inverse shear rate at which the solid is excited. The ratio of these two timescales gives
\begin{equation}
\T \equiv \frac{\tau}{\tau_{\rm p}} = \frac{\tau V}{\ell},
\label{deb}
\end{equation}
the second dimensionless parameter in the problem. $\T$ is the solid analogue of the Deborah number of a viscoelastic fluid. If the material relaxes much faster than the timescale of the changes of its load i.e., $\T\ll 1$, the material behaves purely elastically. If both timescales are comparable ($\T \sim \mathcal{O}(1)$), viscoelasticity becomes important.

It turns out that $\beta$ and $\T$ are the only two dimensionless groups in the problem. This is made explicit by introducing a set of non-dimensional variables,
\begin{equation}
\begin{split}
	&\overbar{x}=\frac{x}{\ell},\hspace{2mm}
	 \overbar{z}=\frac{z}{\Delta},\hspace{2mm}
	 \overbar{h}=\frac{h}{\Delta},\hspace{2mm}
	 \overbar{\mathcal{H}}=\frac{\mathcal{H}}{\mathcal{H}^*},\hspace{2mm}
	 \overbar{p}=\frac{p}{\mathrm{P^*}}\\
	&\overbar{t}=\frac{t}{\tau},\hspace{2mm}
	 %\overbar{\omega}=\omega\tau,\hspace{2mm}
	 \overbar{\mu}=\frac{\mu}{G},\hspace{2mm}
	 \overbar{q}=q\ell,\hspace{2mm}
	 \overbar{L}=\frac{L}{\mathrm{P^*}\ell}.
 \label{ndsplit}
 \end{split}
\end{equation}
In the remainder, we will only use dimensionless quantities and thus drop the overbars.

In dimensionless form \eqref{hreal} becomes
\begin{eqnarray}
\mathcal{H}(x)&=&-\int_{-\infty}^{\infty}\frac{\widetilde{p}(q)}{|q|\mu(-\T q)}\e^{\ii qx}\frac{\id q}{2\upi}\nonumber\\
&=&-\int_{-\infty}^{\infty}\frac{\widetilde{p}(q)}{|q|}\frac{G'(-\T q)-\ii G''(-\T q)}{G'(-\T q)^2+G''(-\T q)^2}\e^{\ii qx}\frac{\id q}{2\upi},
\label{ndhreal}
\end{eqnarray}
which contains $\T$ as a parameter. Likewise, the dimensionless lubrication equation becomes
\begin{equation}
\d{}{x}\left[\d{p}{x}h(x)^3+6h(x)\right]=0.
\label{ndlub}
\end{equation}
The deformed gap profile $h(x)$ couples~\eqref{ndhreal} and~\eqref{ndlub} by 
\begin{equation}
h(x)=h_0(x)-\beta\mathcal{H}(x).
\label{hexp}
\end{equation}
which contains $h_0(x)=1+x^2$ as well as the dimensionless number $\beta$. As anticipated, the problem contains only the two parameters $\T$ and $\beta$. In the case of a purely elastic deformation ($\T=0$), the solution depends only on $\beta$ \citep[]{Bissett1989, Hooke1972, Snoeijer2013}. 

%\textcolor{red}{The comment below is misplaced -- should go to the intro or to the discussion}
%\citet[]{Hooke1997} numerically studied the variations of pressure and film thickness for viscoelastic materials ($\T>0$). Recently, \citet[]{Scaraggi2014} calculated the friction of rolling and sliding lubricated viscoelastic Hertz contacts.
%%%%%%%%%%%%%%%%%%%%%%%%%%%%%%%%%%%%%%%%%%%%%%%%%%%%%%%%%%%%%%%%

\subsection{Solution strategy}

We seek a perturbative solution of~\eqref{ndlub} and~\eqref{hexp} for $\beta\ll1$, in the spirit of previous work on thin compressible elastic layers \citep[]{Skotheim2004}. In this limit, we can expand $p(x)$ in $\beta$ as 

\begin{equation}
p(x)=p_0(x)+\beta p_1(x)+\mathcal{O}(\beta^2).
\end{equation} 

For the leading orders in $\beta$ we get from the lubrication equation~\eqref{ndlub}
\begin{subequations}
\begin{align}
\label{hlub0}
\mathcal{O}(1):\hspace{2mm}&\d{}{x}\left[\d{p_0}{x}h_0^3+6h_0\right]=0,\\
\label{hlub1}
\mathcal{O}(\beta):\hspace{2mm}&\d{}{x}\left[\d{p_1}{x}h_0^3-3h_0^2\mathcal{H}\d{p_0}{x}-6\mathcal{H}\right]=0.
\end{align}
\end{subequations}
Here~\eqref{hlub0} is the steady state classical lubrication problem with rigid boundaries. It can readily be solved with the boundary conditions $p_0(-\infty)=p_0(\infty)=0$:
\begin{equation}
p_0(x)=\frac{2x}{(1+x^2)^2},\hspace{1mm}\widetilde{p}_0(q)=-\ii\upi q\e^{-|q|}.
\label{p0x}
\end{equation}
We note that the zeroth order pressure $p_0(x)$ is antisymmetric and does not contribute to the lift force. Plugging $\widetilde{p}_0(q)$ in~\eqref{ndhreal}, we solve for $\mathcal{H}(x)$ and subsequently solve~\eqref{hlub1} for $p_1(x)$. Then the lift force (per unit length) on the cylinder is
\begin{equation}
 L=\int_{-\infty}^{\infty}p(x)\id x=\beta\int_{-\infty}^{\infty}p_1(x)\id x +\mathcal{O}(\beta^2).
\end{equation} 

It is important to note that the functions $p_1(x)$ and $\mathcal{H}(x)$ depends only on $\T$, so that the lift scales as $L \sim \beta$. The proportionality factor, however, will have a subtle dependence on $\T$; and hence on the lubrication velocity. The primary goal of the analysis will be to identify this $\T$ dependence for different rheological models.

%In the following, we calculate the wall deformation $\mathcal{H}(x)$ from Eq.~\eqref{ndhreal} using $\widetilde{p}_0(q) = \widetilde{p}_0(q)=-\ii\pi q e^{-|q|}$. The result is then used to solve the lubrication equation of linear order in $\beta$~\eqref{hlub1}.

%%%%%%%%%%%%%%%%%%%%%%%%%%%%%%%%%%%%%%%%%%%%%%%%%%%%%%%%%%%%%%%%%%%%%%%%%%%%%%%
\section{Results}\label{sec:results}
We consider viscoelastic lubrication for three different rheological models for the wall, each with one single characteristic timescale: the standard linear solid (SLS), the Kelvin-Voigt model (KV),  and a power law gel (PL). In the following, we first briefly discuss the elastic case and then introduce viscoelasticity through the three different models.
 
\subsection{Elastic wall ($\T$=0)}

For an elastic wall that adopts instantaneously to load changes, the rheology reduces to $\mu=1$. Using~\eqref{ndhreal} and the leading order pressure~(\ref{p0x}), the deformation becomes
\begin{equation}
\mathcal{H}(x)=%\int_{-\infty}^{\infty}\frac{\ii e^{-|q|} q}{2|q|}e^{\ii q x}\id q=
	-\frac{x}{(1+x^2)},
\label{helastic}
\end{equation}
This deformation is purely antisymmetric, just like $p_0(x)$. The first order pressure $p_1$ is obtained by solving~(\ref{hlub1}) using boundary conditions $p_1(-\infty)=p_1(\infty)=0$ to get 
\begin{equation}
p_1(x)=\frac{1-2x^2}{(1+x^2)^4}.
\label{p1elastic}
\end{equation}
The lift force on the cylinder is obtained by integrating over $p_1(x)$ and yields
\begin{equation}
L_0=%\beta\int_{-\infty}^{\infty}\frac{1-2x^2}{(1+x^2)^4}\id x=
	\frac{3\upi}{16}\beta.
\label{lelastic}
\end{equation}
Our main interest is to study how the lift force differs from the purely elastic $L_0$, for various viscoelastic models. The difference will appear when $\T$ is order unity (or larger), for which the viscoelastic solid is excited on timescales comparable to (or faster than) the relaxation time. For $\T \ll 1$ the response will reduce to the elastic case, with $L=L_0$.
%%%%%%%%%%%%%%%%%%%%%%%%%%%%%%%%%%%%%%%%%%%%%%%%%%%%%%%%%%%%%%%%%%%%%%%%%%%%%%
\subsection{Standard Linear Solid Model}

\begin{figure}
	\begin{center}%
		\includegraphics[width=135mm]{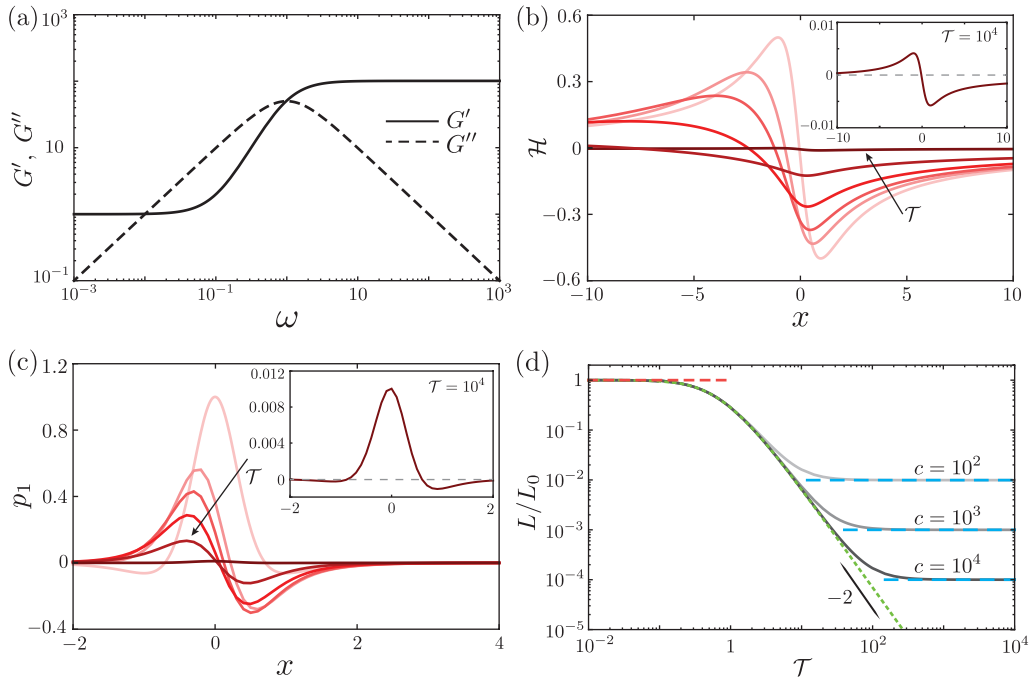}%
	\end{center}%
	\caption{(a) Storage and Loss moduli of SLS for $c=100$. (b) Deformation of SLS half space ($c=100$) due to zeroth order lubrication pressure $p_0$ for $\T=\{0.05, 1, 2, 5, 20, 10^4\}$ (cf.~\eqref{hsls}). Inset: Magnified deformation for $\T=10^4$ shows that it regains small $\T$ shape. (c) First order pressure $p_1$ for $\T=\{0.05, 1, 2, 5, 20, 10^4\}$.  Inset: Magnified pressure at $\T=10^4$. (d) Lift force on the cylinder as a function of $\T$ for three different $c$ values normalized by $L_0$ . The red dashed line is the small $\T$ asymptotic ($L/L_0=1$). The blue dashed lines are the large $\T$ asymptotic given by~\eqref{llargev}. The gray curves represent numerically calculated lift force using the full profile of~\eqref{hsls}. The green dashed curve gives lift force for KV model calculated using~\eqref{hkv}.}
	\label{fig:2}
\end{figure}

The simplest rheological model for a viscoelastic \emph{solid} is represented by a spring and dashpot in parallel, which is the so-called Kelvin-Voigt solid (KV). Here we start the analysis by a considering a generalisation of the KV model, where an additional spring is added in series with the dashpot; this gives the standard linear solid model (SLS). This is three-parameter viscoelastic solid model that has two elastic moduli and a timescale. The two moduli correspond to an instantaneous modulus and a long-time modulus, the former being typically much larger than the latter. For intermediate frequencies, viscous dissipation causes an exponential relaxation behaviour, characterised by a timescale $\tau$. In terms of the dimensionless relaxation function $\Psi(t)$ and complex shear modulus $\mu(\omega)$, the SLS reads 

 \begin{subequations}
 \begin{align}
 &\Psi(t)=1+c \e^{-ct},\\
 &\mu(\omega)=\frac{\omega^2+c^2+c\omega^2}{\omega^2+c^2}+\ii\frac{c^2\omega}{\omega^2+c^2}.
 \label{relaxsls}
 \end{align}
 \end{subequations}
This dimensionless form contains a single parameter: the factor $1/(1+c)$, describing the ratio between static and instantaneous moduli. The storage modulus $G'(\omega)=\Real[\mu(\omega)]$ and loss modulus $G''(\omega)=\Imag[\mu(\omega)]$ are plotted in figure~\ref{fig:2}(a) for $c=100$. One indeed observes two distinct values of $G'$, at low and high frequency respectively. 

For this model, the viscoelastic deformation is obtained in closed form by solving~\eqref{ndhreal}:
 \begin{equation}
 \mathcal{H}(x)=-\frac{x}{(1+c)(1+x^2)}
 +\frac{c^2\e^{\frac{cx}{\T+c\T}}}{\T(1+c)^2}\Real\left[\e^{\frac{\ii c}{\T+c\T}}\left(\mathrm{Ei}\left[-\frac{c(\ii+x)}{\T+c\T}\right]+\ii\upi\right)\right],
 \label{hsls} 
 \end{equation}
where $\mathrm{Ei}$ is the exponential integral of a complex function $z$, defined as $\mathrm{Ei}(z)=-\int_{-z}^{\infty}\frac{\e^{-t}}{t}\id t$. The deformation according to~\eqref{hsls} is plotted in figure~\ref{fig:2}(b). At very small values of $\T$, the response is essentially elastic and $\mathcal{H}$ is perfectly antisymmetric. Viscoelastic effects become apparent for increasing $\T$, for which the deformation decreases in amplitude and loses its perfectly antisymmetric form. However, at very high $\T$, the instantaneous elasticity of the SLS becomes dominant. Hence, for $\T\rightarrow\infty$, one recovers the same profile as for $\T = 0$, but with an amplitude reduced by a factor $1/(1+c)$ (inset figure~\ref{fig:2}(b)).

Using ~\eqref{hsls}, we solve~(\ref{hlub1}) numerically for the first order pressure $p_1$. Integrating $p_1$ we obtain the lift force $L$ on the cylinder. The resulting pressures are shown in figure~\ref{fig:2}(c), while we report the lift force $L/L_0$ (i.e. normalised by the elastic case) in figure~\ref{fig:2}(d). As anticipated, $L/L_0 =1$ in the limit of small $\T$. The lift decreases upon increasing $\T$: the deformation $\mathcal{H}$ drops in amplitude and its shape develops a symmetric component, both leading to a reduction in the generated lift force. At very large $\T$, where the SLS responds instantaneously, the shape and pressure becomes independent of $\T$. As a consequence, the lift force settles at a constant value. Since the effective modulus at short-times is smaller by a factor $1/(1+c)$, the lift force in this limit becomes
\begin{equation}
L/L_0 = (1+c)^{-1},
\label{llargev}
\end{equation}
which is indicated by the blue dashed lines in figure~\ref{fig:2}(d). 

Intriguingly, the numerical results in figure~\ref{fig:2}(d) suggest an intermediate asymptotic regime that emerges when $c \gg 1$, indicated by a dashed green curve. As we will show in the next paragraph, this intermediate regime corresponds exactly to the Kelvin-Voigt (KV) model. It will be shown that for $\T\gg 1$, lift force for KV reads $L\simeq \frac{2}{3}\T^{-2}$. Comparing with~\eqref{llargev}, we see that the intermediate asymptotics is describe by

\begin{equation}
L/L_0=\frac{2}{3}\T^{-2},\hspace{1mm} \mathrm{for}\hspace{1mm}1\ll \T\ll \sqrt{c}.
\label{slsintasymp}
\end{equation}
This regime can indeed be observed when $c \gg 1$, which is naturally expected to be the case for solids that exhibit an instantaneous elasticity.

%%%%%%%%%%%%%%%%%%%%%%%%%%%%%%%%%%%%%%%%%%%%%%%%%%%%%%%%%%%%%%%%%%%%%%%%%%%%%%%
\subsection{Kelvin-Voigt limit}

In the limit of $c\rightarrow\infty$, the instantaneous relaxation of the SLS is suppressed and one recovers the KV model. The relaxation function and the complex modulus of the KV model is given by 
\begin{subequations}
\begin{align}
\Psi(t)=1+\delta(t),\\
\mu(\omega)=1+\ii\omega.
\label{KVcmod}
\end{align}
\end{subequations} 
So for $\omega \ll 1$, KV behaves as purely elastic, while for $\omega \gg 1$ it acts as a Newtonian fluid. The surface deformation of a KV half space is obtained as $c\rightarrow\infty$ in~\eqref{hsls}
\begin{equation}
\mathcal{H}(x;\T)=\frac{\e^{\frac{x}{\T}}}{\T}\Real\left[\e^{\frac{\ii}{\T}}\left(\mathrm{Ei}\left[-\frac{(\ii+x)}{\T}\right]+\ii\upi\right)\right].
\label{hkv}
\end{equation}
The green dashed curve in figure~\ref{fig:2}(d) gives the numerically calculated lift force, which is indeed the intermediate asymptotics of the SLS model. 

We now calculate the asymptotic nature of the lift force for this material model. At large $\T$, ~\eqref{hkv} reduces to
\begin{eqnarray}
\mathcal{H}(x;\T)&=&\T^{-1}(\ugamma+\frac{1}{2}\log(1+x^2)) \nonumber \\
&+&\T^{-2}(x(\ugamma-1)+\frac{x}{2}\log(1+x^2)+\tan^{-1}x) + \mathcal{O}(\T^{-3}),
\label{hkvlargeT}
\end{eqnarray}
where $\ugamma$ is Euler's constant = $0.577216$. Integrating~(\ref{hlub1}), we find
\begin{equation}
p_1=\int_{-\infty}^x 2\left(\frac{1}{h_0}\d{p_0}{x'}+\frac{3}{h_0^3}\right)\mathcal{H}\id x'=\T^{-1}g_1(x)+\T^{-2}g_2(x).
\label{p1int}
\end{equation}
Here $g_1(x)$ is antisymmetric and doesn't contribute to the lift force. The leading order contribution to the lift force is thus $\sim\T^{-2}$: 
\begin{equation}
L=\int_{-\infty}^{\infty}p_1\id x=\T^{-2}\int_{-\infty}^{\infty}g_2(x)\id x=k\T^{-2}.
\label{}
\end{equation} 
Numerical integration over $g_2(x)$ appears to give an exact ratio $k=\pi/8$ up to eight decimals. As a result, the large $\T$ asymptotics for the lift force becomes
\begin{equation}
	L /L_0= \frac{2}{3}\T^{-2},
\label{lkvlarge}
\end{equation} 
which is the scaling law anticipated above.

%%%%%%%%%%%%%%%%%%%%%%%%%%%%%%%%%%%%%%%%%%%%%%%%%%%%%%%%%%%%%%%%%%%%%%%%%%%%%%%%

\subsection{Power Law Rheology}

\begin{figure}
	\begin{center}%
	\includegraphics[width=135mm]{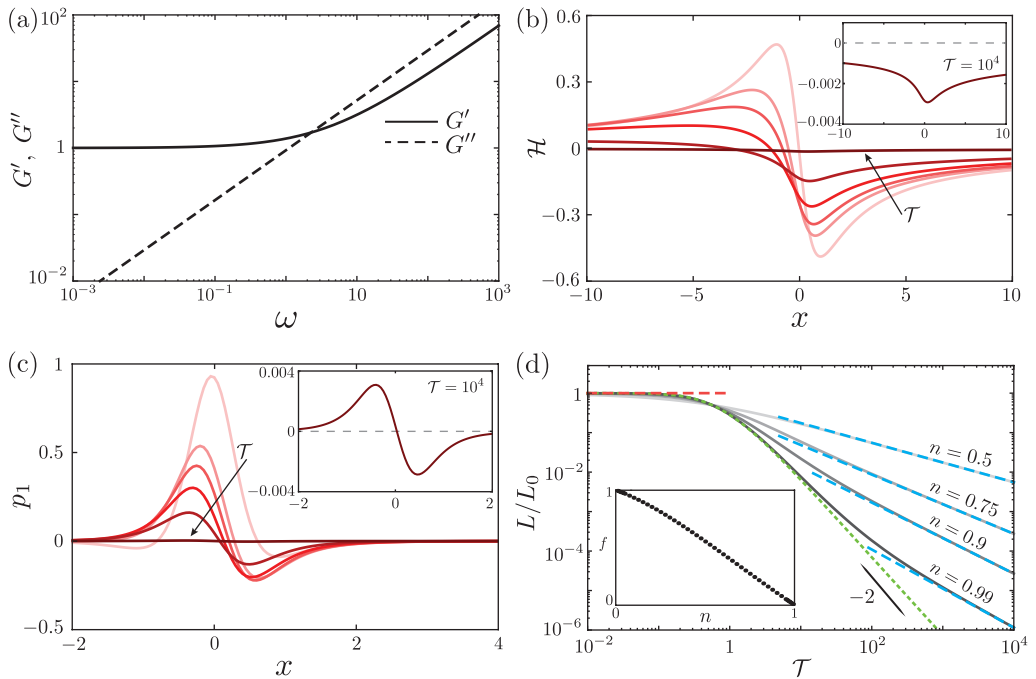}%
	\end{center}%
	\caption{(a) Storage and Loss moduli of PL for $n=3/4$. (b) Deformation of PL half space (with $n=3/4$) obtained by numerical backward transform of~\eqref{plhp} for $\T=\{0.05, 1, 2, 5, 20, 10^4\}$. Inset: Magnified deformation for $\T=10^4$ shows unlike SLS, PL doesn't exhibit elastic deformation for large $\T$. (c) First order pressure $p_1$ for $\T=\{0.05, 1, 2, 5, 20, 10^4\}$. Inset: Magnified pressure profile at $\T=10^4$. (d) Lift force on the cylinder as a function of $\T$ for four different $n$ values. The red dashed line is the small $\T$ approximation given by~\eqref{lelastic}. The blue dashed lines are the large $\T$ asymptotic given by~\eqref{PLlv}. The gray curves represent numerically calculated lift force using the full profile of~\eqref{plhp}. The green dashed curve gives lift force for KV model. Inset: Data points show numerically calculated prefactor $f(n)$ as $n$ varies between 0 and 1.}
	\label{fig:3}
\end{figure}

Many crosslinked polymers like PDMS, Polyurethane exhibit a power-law relaxation behavior at \textit{gel point}, i.e. both $G'(\omega)$ and $G''(\omega)$ scales as $\omega^n$. In general, large degree of polydispersity or a broad relaxation spectrum causes power law relaxation behavior in polymeric systems~\cite[]{Ng08}. The value of the exponent $n$ depends on stoichiometry (ratio of prepolymer to crosslinker). For example, $n=\frac{1}{2}$ for a stoichiometrically balanced PDMS, otherwise $n$ varies between 1/2 and 1~\citep[]{chambon1987linear}. The rheology of such a Power Law gel (PL) can be modelled as:

\begin{subequations}
	\label{psiPL}
 \begin{align}
	 \Psi(t) &= 1+\Gamma(1-n)^{-1}\frac{1}{t^n},\\
	 \mu(\omega) &= 1+(\ii\omega)^n.
 \end{align}
\end{subequations}
where $\Gamma$ is the Gamma function.~\eqref{psiPL} require $0<n<1$, to allow for integrability. Figure~\ref{fig:3}(a) shows the corresponding storage and loss moduli for $n=3/4$. 

The case where $n$ approaches unity is a singular limit, in the sense that the limit of large $\omega$ and $n \rightarrow 1$ cannot be reversed. At a given frequency $\omega$, the response of the PL gel approaches that of the KV rheology in the limit $n \rightarrow 1$, see~(\ref{KVcmod}). However, for any value of $n <1$, the high frequency asymptotics of the PL gel is $G'\sim G''\sim \omega^n$, while for KV one has $G' \sim \omega^0$. For a given material of $n<1$, we therefore anticipate the lift in the high velocity regime ($\T \gg 1$) to be different from the KV behaviour. Again, we will find that the KV model serves as an intermediate regime for the PL solid.  
 
The Fourier transform of the deformation of the PL solid reads
\begin{equation}
 \widetilde{\mathcal{H}}(q)=\frac{\ii \upi \e^{-|q|}q}{|q|(1+(-\ii\T q)^n)}.
 \label{plhp}
\end{equation}
The inversion to real space had to be performed numerically. The results are shown in figure~\ref{fig:3}(b). The inset shows a zoom to a profile for large $\T$. Unlike the SLS model, PL solid does not exhibit an elastic response at large $\T$, and the profile is not perfectly antisymmetric. Figure~\ref{fig:3}(c) shows the corresponding pressure profiles (for $n=0.75$) and lift forces are shown in figure~\ref{fig:3}(d) (gray curves). For small $\T$, the lift force is similar to the elastic case. For large $T$, $L$ decreases algebraically with an exponent that depends on $n$. The green curve corresponds to the KV model.

We now extract the large $\T$ behavior of $L$, and in particular its dependence on the rheological exponent $n$. We expand~\eqref{plhp} and to leading order, get

\begin{equation}
\widetilde{\mathcal{H}}(q)=-\frac{\e^{-|q|}\upi(-\ii q)^{1-n}\T^{-n}}{|q|},
\end{equation}
which can be inverted to real space as

\begin{equation}
	\mathcal{H}(x;\T)=-\Gamma(1-n)\Real\biggl[\frac{\ii^{-n}(1+\ii x)^n}{-\ii+x}\biggr]\T^{-n}.
	\label{plh1x}
\end{equation} 
Using~\eqref{p1int} we can write,
\begin{equation}
	p_1 = \T^{-n} \int_{-\infty}^{x}-3\Gamma(1-n)\Real\biggl[\frac{\ii^{-n}(1+\ii x')^n}{-\ii+x'}\biggr]\left(\frac{1}{h_0}\d{p_0}{x'}+\frac{2}{h_0^3}\right)\id x'=\T^{-n}f_1(x;n),
\end{equation}
and the lift force
\begin{equation}
	L/L_0=\frac{16}{3\upi}\int_{-\infty}^{\infty}\T^{-n}f_1(x;n)  \id x=\T^{-n}f(n).
	\label{PLlv}
\end{equation}
Hence we find that $L \sim \T^{-n}$. The prefactor $f(n)$ is evaluated numerically and shown in the inset of figure~\ref{fig:3}(d). The blue dashed lines in figure~\ref{fig:3}(d) show the large $\T$ asymptotes, in agreement with the full numerical evaluation. 

The KV model again serves as an intermediate regime (green curve), which now emerges as $n \rightarrow 1$. Comparing~\eqref{PLlv} and~\eqref{lkvlarge} we find the scaling 
\begin{equation}
%L/L_0=\frac{2}{3}\T^{-2},\hspace{1mm}\mathrm{for}\hspace{1mm}1\ll \T \ll\left(\frac{3}{2}f(n)\right)^{\frac{1}{n-2}}.
L/L_0=\frac{2}{3}\T^{-2},\hspace{1mm}\mathrm{for}\hspace{1mm}1\ll \T \ll\left(1-n \right)^{-1},
\label{plinterasym}
\end{equation}
where we used that $f \sim (1-n)$ near $n=1$. The upper bound reflects that for $0<1-n\ll 1$, the PL gel does not converge to the KV model at very high frequencies.

%%%%%%%%%%%%%%%%%%%%%%%%%%%%%%%%%%%%%%%%%%%%%%%%%%%%%%%%%%%%%%%%%%%%%%%%%%%%%%%
\section{Discussion}\label{sec:discussion}

We have analysed how the mechanics of lubricated contacts is affected by viscoelastic properties of the solid. We focussed on two-dimensional cylindrical contacts in the limit of small deformations, and considered several different rheological models for the lubricated solid. Here we briefly summarise the key findings, presented in physical units, and discuss how the results are generalised to arbitrary form of $\mu(\omega)= G'(\omega)+\ii G''(\omega)$.

At low lubrication velocities, all rheologies with a non-vanishing static modulus $G=G'(\omega=0)$ give rise to a purely elastic response. In this case, the lift force (per unit length) becomes

\begin{equation}\label{eq:poep}
L_0=\frac{3\pi\eta^2R^2}{8G\Delta^3}V^2.
%L_0=\frac{3\pi\eta^2R^2}{8G\Delta^3}V^2.
\end{equation}
The effect of viscoelasticity is twofold: (i) the resulting deformation is reduced in amplitude with respect to the purely elastic case, and (ii) the deformation profile develops a symmetric part. Both effects give a reduction in the lift force, but the details of this reduction depend on the rheological model. For the Kelvin-Voigt solid, the large velocity asymptote reads

\begin{equation}
L=\frac{\pi\eta^2R^3}{2G\tau^2\Delta^2},
\end{equation}
which interestingly corresponds to a lift that is independent of velocity. For the Power-Law model we find the scaling law

\begin{equation}
L \sim \frac{\eta^2 R^2\ell^n}{G\Delta^3\tau^n}V^{2-n},
\end{equation}
with a prefactor that depends on the rheological exponent $n$. 

It is of interest to generalise these findings to arbitrary rheology. One can identify the relevant scale of the solid deformation upon inspection of~(\ref{ndhreal}), bearing in mind that only the antisymmetric deformation contributes to the lift. From this, we derive that the lift scales as

\begin{equation}
L \sim \frac{\eta^2 R^2}{\Delta^3}\frac{G'(V/\ell)}{G'(V/\ell)^2+G''(V/\ell)^2}V^2,
\end{equation}
which is indeed consistent with all scaling laws mentioned above. Clearly, this implies a reduction of the lift force with respect to the elastic response~(\ref{eq:poep}), whenever the rheology contains a significant contribution of the storage modulus. This expression also highlights the importance of both the storage and a loss modulus to determine the hydrodynamic lift in viscoelastic lubrication: neither a vanishing nor an infinite $G'$ will lead to lift.

The presented formulation may be considered as a rheological tool. Indeed, lubrication has recently been exploited for in-situ AFM measurements of elastic properties of thin films at the microscale~\citep[]{leroy2011}. Including the dissipative nature of the solid into the hydrodynamic interaction force in principle provides access to the full rheological spectrum of squishy layers.

\vspace{10mm}
{\bf Acknowledgments. }
We thank B. Andreotti and T. Salez for discussions. SK acknowledges financial support from NWO through VIDI Grant No. 11304. AP and JS acknowledges financial support from ERC (the European Research Council) Consolidator Grant No. 616918.

\newpage
\bibliography{viscoelastic_lubrication}

\begin{thebibliography}{30}
\expandafter\ifx\csname natexlab\endcsname\relax\def\natexlab#1{#1}\fi
\def\au#1{#1} \def\ed#1{#1} \def\yr#1{#1}\def\at#1{#1}\def\jt#1{\textit{#1}}
  \def\bt#1{#1}\def\bvol#1{\textbf{#1}} \def\vol#1{#1} \def\pg#1{#1}
  \def\publ#1{#1}\def\arxiv#1{#1}\def\org#1{#1}\def\st#1{\textit{#1}}

\bibitem[Bissett(1989)]{Bissett1989}
{\sc \au{Bissett, E.~J.}} \yr{1989}  \at{The line contact problem of
  elastohydrodynamic lubrication. i. asymptotic structure for low speeds}.
  \jt{Proc. R. Soc. A}  \bvol{424}~(1867),  \pg{393--407}.

\bibitem[Bretherton(1961)]{Bretherton1961}
{\sc \au{Bretherton, F.~P.}} \yr{1961}  \at{The motion of long bubbles in
  tubes}.  \jt{J. Fluid Mech.}  \bvol{10},  \pg{166--188}.

\bibitem[Chambon \& Winter(1987)]{chambon1987linear}
{\sc \au{Chambon, F.} \& \au{Winter, H.~H.}} \yr{1987}  \at{Linear
  viscoelasticity at the gel point of a crosslinking {PDMS} with imbalanced
  stoichiometry}.  \jt{J. Rheol.}  \bvol{31}~(8),  \pg{683--697}.

\bibitem[Desrochers {\em et~al.\/}(2012)Desrochers, Amrein \&
  Matyas]{Desrochers2012}
{\sc \au{Desrochers, J.}, \au{Amrein, M.~W.} \& \au{Matyas, J.~R.}} \yr{2012}
  \at{{Viscoelasticity of the articular cartilage surface in early
  osteoarthritis}}.  \jt{Osteoarthr. Cartil.}  \bvol{20}~(5),  \pg{413--421}.

\bibitem[Dowson(1998)]{dowson}
{\sc \au{Dowson, D.}} \yr{1998} {\em History of Tribology\/}.  \publ{Wiley, 2nd
  edition}.

\bibitem[Feng \& Weinbaum(2000)]{Feng2000}
{\sc \au{Feng, J.} \& \au{Weinbaum, S.}} \yr{2000}  \at{Lubrication theory in
  highly compressible porous media: the mechanics of skiing, from red cells to
  humans}.  \jt{J. Fluid Mech.}  \bvol{422},  \pg{281--317}.

\bibitem[Fitz-Gerald(1969)]{Fitz1969}
{\sc \au{Fitz-Gerald, J.~M.}} \yr{1969}  \at{Mechanics of red-cell motion
  through very narrow capillaries}.  \jt{Proc. R. Soc. B}  \bvol{174}~(1035),
  \pg{193--227}.

\bibitem[Hooke \& Huang(1997)]{1997}
{\sc \au{Hooke, C.~J.} \& \au{Huang, P.}} \yr{1997}  \at{{Elastohydrodynamic
  lubrication of soft viscoelastic materials in line contact}}.  \jt{Proc.
  Inst. Mech. Eng. J J. Eng. Tribol.}  \bvol{211}~(3),  \pg{185--194}.

\bibitem[Hooke \& O'Donoghue(1972)]{Hooke1972}
{\sc \au{Hooke, C.~J.} \& \au{O'Donoghue, J.~P.}} \yr{1972}
  \at{Elastohydrodynamic lubrication of soft, highly deformed contacts}.
  \jt{Proc. Inst. Mech. Eng. C J. Mech. Eng. Sci.}  \bvol{14}~(1),
  \pg{34--48}.

\bibitem[Hou {\em et~al.\/}(1992)Hou, Mow, Lai \& Holmes]{Hou1992}
{\sc \au{Hou, J.~S.}, \au{Mow, V.~C.}, \au{Lai, W.M.} \& \au{Holmes, M.H.}}
  \yr{1992}  \at{An analysis of the squeeze-film lubrication mechanism for
  articular cartilage}.  \jt{J. Biomech.}  \bvol{25}~(3),  \pg{247 -- 259}.

\bibitem[Johnson(1987)]{Johnson1987contact}
{\sc \au{Johnson, K.~L.}} \yr{1987} {\em Contact Mechanics\/}.  \publ{Cambridge
  University Press}.

\bibitem[Jones {\em et~al.\/}(2008)Jones, Fulford, Please, McElwain \&
  Collins]{Jones2008}
{\sc \au{Jones, M.~B.}, \au{Fulford, G.~R.}, \au{Please, C.~P.}, \au{McElwain,
  D. L.~S.} \& \au{Collins, M.~J.}} \yr{2008}  \at{{Elastohydrodynamics of the
  eyelid wiper}}.  \jt{Bull. Math. Biol.}  \bvol{70}~(2),  \pg{323--343}.

\bibitem[Karpitschka {\em et~al.\/}(2015)Karpitschka, Das, van Gorcum, Perrin,
  Andreotti \& Snoeijer]{Karpitschka15}
{\sc \au{Karpitschka, S.}, \au{Das, S.}, \au{van Gorcum, M.}, \au{Perrin, H.},
  \au{Andreotti, B.} \& \au{Snoeijer, J.~H.}} \yr{2015}  \at{Droplets move over
  viscoelastic substrates by surfing a ridge}.  \jt{Nat. Commun.}  \bvol{6},
  \pg{7891}.

\bibitem[Leroy \& Charlaix(2011)]{leroy2011}
{\sc \au{Leroy, S.} \& \au{Charlaix, E.}} \yr{2011}  \at{Hydrodynamic
  interactions for the measurement of thin film elastic properties}.  \jt{J.
  Fluid Mech.}  \bvol{674},  \pg{389--407}.

\bibitem[Mani {\em et~al.\/}(2012)Mani, Gopinath \& Mahadevan]{Mani2012}
{\sc \au{Mani, M.}, \au{Gopinath, A.} \& \au{Mahadevan, L.}} \yr{2012}
  \at{{How things get stuck: Kinetics, elastohydrodynamics, and soft
  adhesion}}.  \jt{Phys. Rev. Lett.}  \bvol{108}~(22).

\bibitem[Martin {\em et~al.\/}(2002)Martin, Clain, Buguin \&
  Brochard-Wyart]{Martin2002}
{\sc \au{Martin, A.}, \au{Clain, J.}, \au{Buguin, A.} \& \au{Brochard-Wyart,
  F.}} \yr{2002}  \at{Wetting transitions at soft, sliding interfaces}.
  \jt{Phys. Rev. E}  \bvol{65},  \pg{031605}.

\bibitem[Mow {\em et~al.\/}(1993)Mow, Ateshian \& Spilker]{Mow1993}
{\sc \au{Mow, V.~C.}, \au{Ateshian, G.~A.} \& \au{Spilker, R.~L.}} \yr{1993}
  \at{Biomechanics of diarthrodial joints: A review of twenty years of
  progress}.  \jt{ASME. J Biomech Eng.}  \bvol{115}~(4B),  \pg{460 -- 467}.

\bibitem[Ng \& McKinley(2008)]{Ng08}
{\sc \au{Ng, Trevor S.~K.} \& \au{McKinley, Gareth~H.}} \yr{2008}  \at{Power
  law gels at finite strains: The nonlinear rheology of gluten gels}.  \jt{J.
  Rheol.}  \bvol{52}~(2).

\bibitem[Reynolds(1886)]{Reynolds1886}
{\sc \au{Reynolds, O.}} \yr{1886}  \at{{On the theory of lubrication and its
  application to Mr. Beauchamp Tower's experiments, including an experimental
  determination of the viscosity of olive oil}}.  \jt{Philos. Trans. R. Soc.
  London}  \bvol{177},  \pg{157}.

\bibitem[Salez \& Mahadevan(2015)]{Salez2015}
{\sc \au{Salez, T.} \& \au{Mahadevan, L.}} \yr{2015}  \at{Elastohydrodynamics
  of a sliding, spinning and sedimenting cylinder near a soft wall}.  \jt{J.
  Fluid Mech.}  \bvol{779},  \pg{181--196}.

\bibitem[Scaraggi \& Persson(2014)]{Scaraggi2014}
{\sc \au{Scaraggi, M.} \& \au{Persson, B. N.~J.}} \yr{2014}  \at{{Theory of
  viscoelastic lubrication}}.  \jt{Tribology International}  \bvol{72},
  \pg{118--130}.

\bibitem[Secomb {\em et~al.\/}(1986)Secomb, Skalak, Özkaya \&
  Gross]{Secomb1986}
{\sc \au{Secomb, T.~W.}, \au{Skalak, R.}, \au{Özkaya, N.} \& \au{Gross, J.~F.}}
  \yr{1986}  \at{Flow of axisymmetric red blood cells in narrow capillaries}.
  \jt{J. Fluid Mech.}  \bvol{163},  \pg{405--423}.

\bibitem[Sekimoto \& Leibler(1993)]{Sekimoto1993}
{\sc \au{Sekimoto, K.} \& \au{Leibler, L.}} \yr{1993}  \at{{A Mechanism for
  Shear Thickening of Polymer-Bearing Surfaces : Elasto-Hydrodynamic Coupling
  .}}  \jt{Europhys. Lett.}  \bvol{23}~(2),  \pg{113--117}.

\bibitem[Skotheim \& Mahadevan(2004)]{Skotheim2004}
{\sc \au{Skotheim, J.~M.} \& \au{Mahadevan, L.}} \yr{2004}  \at{{Soft
  lubrication}}.  \jt{Phys. Rev. Lett.}  \bvol{92}~(24),  \pg{245509--1}.

\bibitem[Skotheim \& Mahadevan(2005)]{Skotheim2005a}
{\sc \au{Skotheim, J.~M.} \& \au{Mahadevan, L.}} \yr{2005}  \at{{Soft
  lubrication: The elastohydrodynamics of nonconforming and conforming
  contacts}}.  \jt{Phys. Fluids}  \bvol{17}~(9),  \pg{1--23}.

\bibitem[Snoeijer {\em et~al.\/}(2013)Snoeijer, Eggers \& Venner]{Snoeijer2013}
{\sc \au{Snoeijer, J.~H.}, \au{Eggers, J.} \& \au{Venner, C.~H.}} \yr{2013}
  \at{{Similarity theory of lubricated Hertzian contacts}}.  \jt{Phys. Fluids}
  \bvol{25}~(10).

\bibitem[Snoeijer \& van~der Weele(2014)]{SnoeijerAJP14}
{\sc \au{Snoeijer, J.~H.} \& \au{van~der Weele, K.}} \yr{2014}  \at{Physics of
  the granite sphere fountain}.  \jt{Am. J. Phys.}  \bvol{82}~(11).

\bibitem[Trickey {\em et~al.\/}(2000)Trickey, Lee \& Guilak]{Trickery2000}
{\sc \au{Trickey, W.~R.}, \au{Lee, G.~M.} \& \au{Guilak, F.}} \yr{2000}
  \at{Viscoelastic properties of chondrocytes from normal and osteoarthritic
  human cartilage}.  \jt{J. Orthop. Res.}  \bvol{18}~(6),  \pg{891--898}.

\bibitem[Urzay {\em et~al.\/}(2007)Urzay, {Llewellyn Smith} \&
  Glover]{Urzay2007}
{\sc \au{Urzay, J.}, \au{{Llewellyn Smith}, S.~G.} \& \au{Glover, B.~J.}}
  \yr{2007}  \at{{The elastohydrodynamic force on a sphere near a soft wall}}.
  \jt{Phys. Fluids}  \bvol{19}~(10).

\bibitem[Venner \& Lubrecht(2000)]{venner2000multi}
{\sc \au{Venner, C.~H.} \& \au{Lubrecht, A.~A.}} \yr{2000} {\em Multi-Level
  Methods in Lubrication\/}.  \publ{Elsevier Science}.

\end{thebibliography}
\bibliographystyle{jfm}

\end{document}